\documentclass[nofootinbib,preprint,preprintnumbers,a4paper,10pt]{revtex4}
\usepackage{multirow}
\usepackage{textcomp}
\usepackage{amsmath,graphicx,color,epsfig}

\usepackage{footnote}
\usepackage{ulem}
\usepackage{booktabs}
\usepackage{array}
\usepackage{amssymb}
\usepackage{subfigure}
\usepackage{longtable}
\usepackage{verbatim}
\usepackage{amsfonts}
\usepackage{hyperref}
\usepackage{cancel}

\begin{document}

\title{$U$-spin sum rules in charmed baryon decays beyond the $U$-spin limit}

\author{Si-Jia Wen$^{1}$}
\author{Di Wang$^{1}$}\email{wangdi@hunnu.edu.cn}

\address{%
$^1$Department of Physics, Hunan Normal University, Changsha, 410081, China
}

\begin{abstract}
$U$-spin symmetry is an important flavor symmetry in studying charmed hadron weak decays.
However, the sizeable $U$-spin breaking limits the predictive power of $U$-spin symmetry.
In this work, we derive dozens of rate and decay parameter $U$-spin sum rules for charmed baryon decays beyond the exact $U$-spin symmetry limit.
These sum rules provide tests of flavor symmetry up to the first-order $U$-spin breaking in the charmed baryon sector.

\end{abstract}

\maketitle


\section{Introduction}
Charmed baryon decays provide laboratories to study strong and weak interactions in heavy-to-light baryonic transitions.
In theoretical aspect, the quantum chromodynamics (QCD)-inspired approaches do not work well in the charmed hadron decays because the expansion parameters $\alpha_s(m_c)$ and $\Lambda_{\rm QCD}/m_c$ are larger than those in the $b$-hadron decays.
Apart from model calculations, it is useful to study charmed baryon decays based on the flavor symmetry.
In literature, the two-body nonleptonic decays of charmed baryons are widely studied in the flavor $SU(3)$ symmetry \cite{Wang:2022wrb,Lai:2025ryi,He:2024unv,Liu:2023dvg,Wang:2025uie,Wang:2023uea,Hsiao:2021nsc,Hsiao:2020iwc,
Hsiao:2019yur,Zhao:2018mov,Geng:2023pkr,Geng:2020zgr,Geng:2019awr,
Geng:2018rse,Geng:2018plk,Geng:2017mxn,Geng:2017esc,Huang:2021aqu,
Wu:2026cyf,Cheng:2025oyr,Cheng:2024lsn,Zhong:2024qqs,He:2018joe,
Wang:2017azm,Lu:2016ogy,Lai:2026xst,Liu:2025hbf,Wang:2024nxb,
Wang:2024ztg,Wang:2022kwe,Jia:2019zxi,Wang:2017gxe} and its $SU(2)$ sub-group, isospin \cite{Luo:2023vbx} and $U$-spin \cite{Gavrilova:2026ufq,Wang:2025ucg,Wang:2019dls,Geng:2026qbv,Gavrilova:2022hbx} symmetries.
Recent studies have attempted to include the flavor symmetry breaking effects \cite{Gavrilova:2026ryc,Geng:2018bow,Geng:2025wfi,Yang:2025orn}.
$U$-spin symmetry reflects the symmetry between the $d$ and $s$ quarks, which is more powerful than isospin and $V$-spin symmetries.
$U$-spin symmetry can connect the Cabibbo-favored (CF) and the doubly Cabibbo-suppressed (DCS) modes, as well as the $c\to s\overline su$ and $c\to d\overline du$ transitions in the singly Cabibbo-suppressed (SCS) modes.
However, isospin and $V$-spin symmetries can only connect one type of charmed hadron decays, as the $d$ and $s$ quarks belong to different isospin and $V$-spin multiplets.
The $d$ and $s$ quarks in the loop diagrams allows $U$-spin to relate the direct $CP$ asymmetries of charmed hadron decays.

$U$-spin breaking is naively estimated to be $\epsilon_U\sim (m_s-m_d)/\Lambda_{\rm had}\sim (20\sim$ $30)\%$, where $\Lambda_{\rm had}$ is the typical strong-interaction scale with $(m_s-m_d)\approx 90$\,MeV and $\Lambda_{\rm had}\approx (300\sim 500)$\,MeV.
Theoretical predictions based on $U$-spin symmetry exhibit deviations from the true values.
To obtain more accurate theoretical predictions, it is necessary to study $U$-spin sum rules that go beyond the $U$-spin limit, i.e., the exact $U$-spin symmetry, as in $D$ meson decays \cite{Muller:2015lua,Muller:2015rna,Grossman:2012ry}.
In Refs.~\cite{Brod:2012ud,Gronau:2013xba,Gronau:2015rda,Gronau:2013mda,Feldmann:2012js,Hiller:2012xm,Pirtskhalava:2011va}, a perturbative method for analyzing $U$-spin breaking was proposed.
In this method, $U$-spin breaking corrections to arbitrary order are obtained by introducing an $s$-$d$ spurion mass operator $M_{\rm Ubrk}$ into the effective Hamiltonian or into the initial and final states.
$M_{\rm Ubrk}$ is the $U_3=0$ component of a $U$-spin triplet.
Using $M_{\rm Ubrk}$, one can derive $U$-spin sum rules beyond the $U$-spin limit.
In this work, we apply this approach to the charmed baryon decays.
Dozens of rate and decay parameter $U$-spin sum rules for charmed baryon decays beyond the $U$-spin limit are derived, providing precise tests of flavor symmetry.

The rest of this paper is structured as follows.
In Sec. \ref{TH}, we take a $U$-spin system as an example to derive rate and decay parameter sum rules beyond the $U$-spin limit.
The other sum rules for singly and doubly charmed baryon decays are presented in Sec. \ref{U}.
Sec. \ref{summary} provides a short summary.

\section{Theoretical framework}\label{TH}

In $U$-spin symmetry, the $d$ and $s$ quarks form a $U=1/2$ doublet.
The $(d,s)$-quark mass term can be written as
\begin{align}
  \mathcal{L}_m^{(d,s)} = -\frac{m_s+m_d}{2} \,(\overline dd+\overline ss) - \frac{m_s-m_d}{2}\,(\overline ss-\overline dd).
\end{align}
The first term is a $U$-spin scalar, and the second term transforms as the $U_3=0$ component of a $U$-spin triplet and is represented by the spurion operator $M_{\rm Ubrk}$ \cite{Gronau:2013xba}.
At the hadronic level, $U$-spin breaking is encoded not only in the reduced weak matrix elements but also in hadron-mass splittings, phase-space and partial-wave momentum factors, decay constants, form factors, and final-state rescattering effects.
Consequently, the effective $U$-spin-breaking parameters are generally complex and depend on the decay system and partial wave.
The dimensionless $U$-spin-breaking expansion parameter $\epsilon_U$ can be naively predicted as
\begin{align}
 \epsilon_U \sim \frac{m_s-m_d}{\Lambda_{\rm had}}\approx \frac{90\,\rm MeV}{(300\sim 500)\,\rm MeV}\approx(20\sim 30)\%.
\end{align}
An intuitive example of $U$-spin breaking is the splitting of the kaon and pion decay constants, i.e., $\epsilon_U\sim f_K/f_{\pi}-1\approx 0.2$.

Taking the $U$-spin system
\begin{align}\label{f4}
    \begin{array}{c}
      \Xi^0_c
    \end{array}
  \to
    \left(
    \begin{array}{c}
      K^+ \\
      \pi^+ \\
    \end{array}
  \right)
    \left(
    \begin{array}{c}
      \Sigma^- \\
      \Xi^- \\
    \end{array}
  \right)
\end{align}
as an example, we illustrate the derivation of $U$-spin amplitudes beyond the $U$-spin limit.
There are four decay channels in this $U$-spin system, $\Xi^0_c\to \Xi^-\pi^+$, $\Xi^0_c\to \Sigma^-K^+$, $\Xi^0_c\to \Xi^-K^+$, and $\Xi^0_c\to \Sigma^-\pi^+$.
The decays $\Xi^0_c\to \Xi^-\pi^+$ and $\Xi^0_c\to \Sigma^-K^+$, as well as $\Xi^0_c\to \Xi^-K^+$ and $\Xi^0_c\to \Sigma^-\pi^+$, are two $U$-spin conjugate pairs, i.e., the two decay modes are related by the interchanges of $s\leftrightarrow d$ and $\overline s\leftrightarrow -\overline d$ in all the initial and final states.
If two decay channels are $U$-spin conjugate pairs, the third $U$-spin components of the corresponding initial and final states have equal in magnitude but opposite in sign.
For example, the $U$-spins and corresponding third components of the $\Xi^0_c$, $K^+$, $\pi^+$, $\Sigma^-$, and $\Xi^-$ are $(0,0)$, $(1/2,1/2)$, $(1/2,-1/2)$, $(1/2,1/2)$, and $(1/2,-1/2)$, respectively.
The three pieces of the Hamiltonian responsible for CF, SCS, and DCS modes behaving like $\overline sd $, ($s \overline s - d \overline d$), and $\overline ds $ form a $U$-spin triplet \cite{Gronau:2013xba},
\begin{align}
  \mathcal{H}_{\rm CF}=-\cos^2\theta_C\,(1,-1),\qquad \mathcal{H}_{\rm SCS}=\sqrt{2}\,\cos\theta_C\sin\theta_C\,(1,0),\qquad\mathcal{H}_{\rm DCS}=-\sin^2\theta_C\,(1,1),
\end{align}
where $\theta_C$ is the Cabibbo angle.
In the $U$-spin limit, the $U$-spin amplitudes for these four decay modes are given by
\begin{align}
\mathcal{A}(\Xi^0_c\to \Xi^-\pi^+) &= \langle \Xi^-\pi^+|\mathcal{H}_{\rm CF}|\Xi^0_c\rangle= -\cos^2\theta_C\langle \frac{1}{2},-\frac{1}{2}; \frac{1}{2},-\frac{1}{2}|1,-1; 0,0\rangle=-\cos^2\theta_C\mathcal{A}_1,\nonumber\\
\mathcal{A}(\Xi^0_c\to \Sigma^-K^+) &= \langle \Sigma^-K^+|\mathcal{H}_{\rm DCS}|\Xi^0_c\rangle= -\sin^2\theta_C\,\langle \frac{1}{2},\frac{1}{2}; \frac{1}{2},\frac{1}{2}|1,1; 0,0\rangle= -\sin^2\theta_C\,\mathcal{A}_1,\nonumber\\
\mathcal{A}(\Xi^0_c\to \Xi^-K^+) &= \langle \Xi^-K^+|\mathcal{H}_{\rm SCS}|\Xi^0_c\rangle= \sqrt{2}\cos\theta_C\sin\theta_C\,\langle \frac{1}{2},-\frac{1}{2}; \frac{1}{2},\frac{1}{2}|1,0; 0,0\rangle=\cos\theta_C\sin\theta_C\,\mathcal{A}_1,\nonumber\\
\mathcal{A}(\Xi^0_c\to \Sigma^-\pi^+) &= \langle \Sigma^-\pi^+|\mathcal{H}_{\rm SCS}|\Xi^0_c\rangle= \sqrt{2}\cos\theta_C\sin\theta_C\,\langle \frac{1}{2},\frac{1}{2}; \frac{1}{2},-\frac{1}{2}|1,0; 0,0\rangle=\cos\theta_C\sin\theta_C\,\mathcal{A}_1.
\end{align}
Then we get the amplitude relations in the $U$-spin limit as follows:
\begin{align}\label{s4}
 -\mathcal{A}(\Xi^0_c\to \Xi^-\pi^+) = -\frac{\mathcal{A}(\Xi^0_c\to \Sigma^-K^+)}{\tan^2\theta_C}= \frac{\mathcal{A}(\Xi^0_c\to \Xi^-K^+)}{\tan\theta_C}=\frac{\mathcal{A}(\Xi^0_c\to \Sigma^-\pi^+)}{\tan\theta_C} = \mathcal{A}_{1}.
\end{align}

Considering first-order $U$-spin breaking, an $s-d$ spurion mass operator $M_{\rm Ubrk}$, $M_{\rm Ubrk}\propto (s \overline s - d \overline d)=\sqrt{2}(1,0)$, couples to the initial and final states as well as effective Hamiltonian.
According to the angular momentum coupling rule, a simultaneous reversal of all magnetic quantum numbers changes the coupling
coefficients only by the overall phase \cite{Varshalovich:1988ifq},
\begin{align}\label{s3}
 \langle j_1,-m_1;j_2,-m_2|j_{3},-m_{3};j_{4},-m_{4}\rangle
=(-1)^{j_1+j_2-j_{3}-j_4}\langle &j_1,m_1;j_2,m_2|j_{3},m_{3};j_{4},m_{4}\rangle.
\end{align}
Here, $j_i$ denotes the magnitude of the corresponding angular momentum, while $m_i$ denotes its projection along $z$ axis.
The detailed derivation of Eq.~\eqref{s3} is presented in Appendix~\ref{amcr}.
According to Eq.~\eqref{s3}, the first-order $U$-spin breaking corrections for the four decay modes satisfy
\begin{align}
 \mathcal{A}_{\rm Ubrk}(\Xi^0_c\to \Xi^-\pi^+) &=  \frac{(-1)^{|\Delta U|}\mathcal{A}_{\rm Ubrk}(\Xi^0_c\to \Sigma^-K^+)}{\tan^2\theta_C}, \nonumber\\
  \mathcal{A}_{\rm Ubrk}(\Xi^0_c\to \Sigma^-\pi^+) &=  (-1)^{|\Delta U|}\mathcal{A}_{\rm Ubrk}(\Xi^0_c\to \Xi^-K^+),
\end{align}
where $\Delta U=U_\mathcal{B}+U_M-U_{\mathcal{B}_{c}}-U_{\mathcal{H}_{\rm eff}}\pm U_{\rm Ubrk}$ and $U_\mathcal{B}$, $U_M$, $U_{\mathcal{B}_{c}}$, $U_{\mathcal{H}_{\rm eff}}$, and $U_{\rm Ubrk}$ are the $U$-spin quantum numbers of the final state baryon $\mathcal{B}$, final state meson $M$, initial state baryon $\mathcal{B}_{c}$, effective Hamiltonian $\mathcal{H}_{\rm eff}$, and $s-d$ spurion mass operator $M_{\rm Ubrk}$, respectively.
The sign before $U_{\rm Ubrk}$ is "$+$" if $M_{\rm Ubrk}$ couples to one of the final states, and "$-$" if the $M_{\rm Ubrk}$ is coupling with initial state or effective Hamiltonian.
Substituting the $U$-spin quantum numbers of initial/final states, effective Hamiltonian, and $M_{\rm Ubrk}$, we obtain $|\Delta U| =1$.
Then, we have
\begin{align}
 \mathcal{A}_{\rm Ubrk}(\Xi^0_c\to \Xi^-\pi^+)&=-\frac{\mathcal{A}_{\rm Ubrk}(\Xi^0_c\to \Sigma^-K^+)}{\tan^2\theta_C} =  \mathcal{A}_{1}\times \epsilon_1, \nonumber\\
  \frac{\mathcal{A}_{\rm Ubrk}(\Xi^0_c\to \Sigma^-\pi^+)}{\tan\theta_C} &=  -\frac{\mathcal{A}_{\rm Ubrk}(\Xi^0_c\to \Xi^-K^+)}{\tan\theta_C}=\mathcal{A}_{1}\times \epsilon_2,
\end{align}
where $\epsilon_1$ and $\epsilon_2$ are complex $U$-spin breaking parameters.

More generally, $U$-spin breaking of order $n$ can be obtained by introducing in
the Hamiltonian or in the initial/final state a total of $n$ powers of the $s-d$ spurion mass operator $M_{\rm Ubrk}$.
The $n$th-order $U$-spin breaking corrections for the four decay modes satisfy
\begin{align}
 \mathcal{A}^n_{\rm Ubrk}(\Xi^0_c\to \Xi^-\pi^+) &=  \frac{(-1)^{|\Delta U^n|}\mathcal{A}_{\rm Ubrk}(\Xi^0_c\to \Sigma^-K^+)}{\tan^2\theta_C}, \nonumber\\
  \mathcal{A}^n_{\rm Ubrk}(\Xi^0_c\to \Sigma^-\pi^+) &=  (-1)^{|\Delta U^n|}\mathcal{A}^n_{\rm Ubrk}(\Xi^0_c\to \Xi^-K^+),
\end{align}
where $\Delta U^n=U_\mathcal{B}+U_M-U_{\mathcal{B}_{b}}-U_{\mathcal{H}_{\rm eff}}+\sum_1^n (\pm U_{\rm Ubrk})$.
According to the angular momentum coupling rule, $\Delta U^n$ is odd if $n$ is odd, and $\Delta U^n$ is even if $n$ is even.
As a result, the $U$-spin amplitudes for the $\Xi^0_c\to \Xi^-\pi^+$, $\Xi^0_c\to \Sigma^-K^+$, $\Xi^0_c\to \Xi^-K^+$, and $\Xi^0_c\to \Sigma^-\pi^+$ modes at arbitrary order are
\begin{align}\label{s6}
 \mathcal{A}(\Xi^0_c\to \Xi^-\pi^+) &= -\cos^2\theta_C\mathcal{A}_{1}(1- \epsilon_1+a_1\epsilon^2_1-a^\prime_1\epsilon^3_1+...),\nonumber\\
\mathcal{A}(\Xi^0_c\to \Sigma^-K^+) &= -\sin^2\theta_C\mathcal{A}_{1}(1+ \epsilon_1+a_1\epsilon^2_1+a^\prime_1\epsilon^3_1+...),\nonumber\\
 \mathcal{A}(\Xi^0_c\to \Xi^-K^+)&= \cos\theta_C\sin\theta_C\,\mathcal{A}_{1}(1- \epsilon_2+a_2\epsilon^2_2-a^\prime_2\epsilon^3_2+...),\nonumber\\ \mathcal{A}(\Xi^0_c\to \Sigma^-\pi^+)&= \cos\theta_C\sin\theta_C\,\mathcal{A}_{1}(1+ \epsilon_2+a_2\epsilon^2_2+a^\prime_2\epsilon^3_2+...).
\end{align}
The nonperturbative coefficients $a_{1,2}$, $a^\prime_{1,2}$ ..., could be determined from nonperturbative QCD calculations or extracted from experimental data.

The decay width $\Gamma$ is
\begin{align}\label{s15}
&\Gamma = \frac{p_c}{8\pi}\frac{(m_{\mathcal{B}_c}+m_\mathcal{B})^2-m_M^2}
{m_{\mathcal{B}_c}^2}\left(|S|^2
+ |P|^2\right),
\end{align}
where $S$ and $P$ are the $S$-wave and $P$-wave amplitudes, and $p_c$ is the center of momentum (CM) in the rest frame of the initial baryon.
The $U$-spin amplitudes given by Eq.~\eqref{s6} hold for all partial-wave amplitudes.
According to Eq.~\eqref{s6}, we derive a rate sum rule protected against first-order $U$-spin breaking:
\begin{align}\label{s7}
  |\mathcal{A}(\Xi^0_c\to \Xi^-\pi^+)|^2+\frac{|\mathcal{A}(\Xi^0_c\to \Sigma^-K^+)|^2}{\tan^4\theta_C}=
  \frac{|\mathcal{A}(\Xi^0_c\to \Xi^-K^+)|^2}{\tan^2\theta_C} +\frac{|\mathcal{A}(\Xi^0_c\to \Sigma^-\pi^+)|^2}{\tan^2\theta_C} +  \mathcal{O}(\epsilon^2).
\end{align}
The modulus squared of the amplitude $|\mathcal{A}(\mathcal{B}_c\to \mathcal{B}M)|^2$ is calculated from the $S$-wave and $P$-wave amplitudes as $|S(\mathcal{B}_c\to \mathcal{B}M)|^2+ |P(\mathcal{B}_c\to \mathcal{B}M)|^2$.
According to Eq.~\eqref{s15}, $|\mathcal{A}(\mathcal{B}_c\to \mathcal{B}M)|^2$ can be extracted from the branching fraction of $\mathcal{B}_c\to \mathcal{B}M$ mode,
\begin{align}
|\mathcal{A}(\mathcal{B}_c\to \mathcal{B}M)|^2=\frac{8\pi}{p_c\,\tau_{\mathcal{B}_c}}
\frac{m_{\mathcal{B}_c}^2\,\mathcal{B}r(\mathcal{B}_c\to \mathcal{B}M)}
{(m_{\mathcal{B}_c}+m_\mathcal{B})^2-m_M^2}.
\end{align}
The LHS and RHS of Eq.~\eqref{s7} can also be replaced by the product of modulus of two decay amplitudes,
\begin{align}\label{s12}
 \frac{|\mathcal{A}(\Xi^0_c\to \Xi^-\pi^+)|\,|\mathcal{A}(\Xi^0_c\to \Sigma^-K^+)|}{\tan^2\theta_C}
\end{align}
and
\begin{align}\label{s13}
\frac{|\mathcal{A}(\Xi^0_c\to \Xi^-K^+)|\, |\mathcal{A}(\Xi^0_c\to \Sigma^-\pi^+)|}{\tan^2\theta_C}
\end{align}
respectively, and the modulus of the amplitude $|\mathcal{A}(\mathcal{B}_c\to \mathcal{B}M)|$ is calculated from the $S$-wave and $P$-wave amplitudes as $\sqrt{|S(\mathcal{B}_c\to \mathcal{B}M)|^2+ |P(\mathcal{B}_c\to \mathcal{B}M)|^2}$.

The Lee-Yang parameters $\alpha$, $\beta$, and $\gamma$ are defined as
\begin{align}
\alpha=\frac{2\mathcal{R}e (S^*P)}{|S|^2+ |P|^2},\qquad
\beta=\frac{2 \mathcal{I}m(S^*P)}{|S|^2+|P|^2},\qquad
\gamma=\frac{|S|^2- |P|^2}{|S|^2+ |P|^2}.
\end{align}
The amplitude decompositions listed in Eq.~\eqref{s6} apply to both the $S$-wave and $P$-wave amplitudes.
The decay parameter $\alpha$ for $\Xi^0_c\to \Xi^-\pi^+$ is given by
  \begin{align}
 \alpha(\Xi^0_c\to \Xi^-\pi^+) &= \frac{2\mathcal{R}e(S^*P)}{|S|^2+|P|^2} =
 \frac{2\mathcal{R}e[(\mathcal{A}^S_{1/2}(1- \epsilon_1^S...))^*\times(\mathcal{A}^P_{1/2}(1- \epsilon_1^P...))]}{|\mathcal{A}^S_{1/2}(1- \epsilon_1^S...)|^2+|\mathcal{A}^P_{1/2}(1- \epsilon_1^P...)|^2}
 \nonumber\\ & \qquad\simeq  \frac{1}{|\mathcal{A}^S_{1/2}|^2+|\mathcal{A}^P_{1/2}|^2}\times
 \big[2\mathcal{R}e(\mathcal{A}^{S*}_{1/2}\mathcal{A}^{P}_{1/2})-
 2\mathcal{R}e\big((\mathcal{A}^{S}_{1/2}\epsilon_1^{S})^*\mathcal{A}^{P}_{1/2}
 +\mathcal{A}^{S*}_{1/2}\mathcal{A}^{P}_{1/2}\epsilon_1^{P}\big)\nonumber\\&
 \qquad\qquad +2|\mathcal{A}^S_{1/2}|^2\mathcal{R}e( \epsilon_1^{S})+2|\mathcal{A}^P_{1/2}|^2\mathcal{R}e (\epsilon_1^{P})\big]+\mathcal{O}(\epsilon^2_1).
\end{align}
The corresponding decay parameter for $\Xi^0_c\to \Sigma^-K^+$ is
  \begin{align}
 \alpha(\Xi^0_c\to \Sigma^-K^+) & \simeq  \frac{1}{|\mathcal{A}^S_{1/2}|^2+|\mathcal{A}^P_{1/2}|^2}\times
 \big[2\mathcal{R}e(\mathcal{A}^{S*}_{1/2}\mathcal{A}^{P}_{1/2})+
 2\mathcal{R}e((\mathcal{A}^{S}_{1/2}\epsilon_1^{S})^*\mathcal{A}^{P}_{1/2}
 +\mathcal{A}^{S*}_{1/2}\mathcal{A}^{P}_{1/2}\epsilon_1^{P})\nonumber\\&
 \qquad -2|\mathcal{A}^S_{1/2}|^2\mathcal{R}e (\epsilon_1^{S})-2|\mathcal{A}^P_{1/2}|^2\mathcal{R}e (\epsilon_1^{P})\big]+\mathcal{O}(\epsilon^2_1).
\end{align}
Summing $\alpha(\Xi^0_c\to \Xi^-\pi^+)$ and $\alpha(\Xi^0_c\to \Sigma^-K^+)$, we get
  \begin{align}
 \alpha(\Xi^0_c\to \Xi^-\pi^+)+\alpha(\Xi^0_c\to \Sigma^-K^+)\simeq  \frac{4\mathcal{R}e(\mathcal{A}^{S*}_{1/2}\mathcal{A}^{P}_{1/2})}
 {|\mathcal{A}^S_{1/2}|^2+|\mathcal{A}^P_{1/2}|^2}+\mathcal{O}(\epsilon^2_1).
\end{align}
Similarly, we have
\begin{align}
 \alpha(\Xi^0_c\to \Xi^-K^+)+\alpha(\Xi^0_c\to \Sigma^-\pi^+)\simeq  \frac{4\mathcal{R}e(\mathcal{A}^{S*}_{1/2}\mathcal{A}^{P}_{1/2})}
 {|\mathcal{A}^S_{1/2}|^2+|\mathcal{A}^P_{1/2}|^2}+\mathcal{O}(\epsilon^2_2).
\end{align}
We therefore derive the following decay-parameter sum rule, which is protected against first-order U-spin-breaking corrections:
\begin{align}\label{s17}
 \alpha(\Xi^0_c\to \Xi^-\pi^+) +\alpha(\Xi^0_c\to \Sigma^-K^+) =
 \alpha(\Xi^0_c\to \Xi^-K^+)+\alpha(\Xi^0_c\to \Sigma^-\pi^+) + \mathcal{O}(\epsilon^2),
\end{align}
Similarly, the following $U$-spin relations for the decay parameters $\beta$ and $\gamma$ can be derived:
\begin{align}
 \beta(\Xi^0_c\to \Xi^-\pi^+) +\beta(\Xi^0_c\to \Sigma^-K^+) =
 \beta(\Xi^0_c\to \Xi^-K^+)+\beta(\Xi^0_c\to \Sigma^-\pi^+) + \mathcal{O}(\epsilon^2),
\end{align}
\begin{align}\label{s18}
 \gamma(\Xi^0_c\to \Xi^-\pi^+) +\gamma(\Xi^0_c\to \Sigma^-K^+) =
 \gamma(\Xi^0_c\to \Xi^-K^+)+\gamma(\Xi^0_c\to \Sigma^-\pi^+) + \mathcal{O}(\epsilon^2).
\end{align}

Owing to the limited experimental data currently available, the complex $U$-spin breaking parameters $\epsilon_1^S$, $\epsilon_1^P$, $\epsilon_2^S$, and $\epsilon_2^P$ cannot yet be determined so far.
The combination of them can be roughly estimated from the ratio of the branching fractions $\mathcal{B}r(\Xi^0_c\to \Xi^-K^+)$ and $\mathcal{B}r(\Xi^0_c\to \Xi^-\pi^+)$:
\begin{align}
&\frac{|S|^2\big(\mathcal{R}e(\epsilon_1^S)
-\mathcal{R}e(\epsilon_2^S)\big)+|P|^2\big(\mathcal{R}e(\epsilon_1^P)
-\mathcal{R}e(\epsilon_2^P)\big)}{|S|^2+|P|^2} \simeq \left[\frac{|\mathcal{A}(\Xi^0_c\to \Xi^-K^+)|^2}{\tan^2\theta_C\,|\mathcal{A}(\Xi^0_c\to \Xi^-\pi^+)|^2}-1\right]/2 + \mathcal{O}(\epsilon^2)\nonumber\\
&=\left[\frac{\mathcal{B}r(\Xi^0_c\to \Xi^-K^+)}{\tan^2\theta_C\,\mathcal{B}r(\Xi^0_c\to \Xi^-\pi^+)}\frac{\kappa(\Xi^0_c\to \Xi^-\pi^+)}{\kappa(\Xi^0_c\to \Xi^-K^+)}-1\right]/2 + \mathcal{O}(\epsilon^2),
\end{align}
where $\kappa=p_c[(m_{\mathcal{B}_c}+m_{\mathcal{B}})^2-m_M^2]$.
With the ratio $\mathcal{B}r(\Xi^0_c\to \Xi^-K^+)/\mathcal{B}r(\Xi^0_c\to \Xi^-\pi^+)=0.0275\pm0.0057$ given by Belle Collaboration \cite{Belle:2013ntc},
the parameter combination is estimated as
\begin{align}
  \frac{|S|^2\big(\mathcal{R}e(\epsilon_1^S)
-\mathcal{R}e(\epsilon_2^S)\big)+|P|^2\big(\mathcal{R}e(\epsilon_1^P)
-\mathcal{R}e(\epsilon_2^P)\big)}{|S|^2+|P|^2} \approx -0.21\pm 0.06,
\end{align}
which is consistent with the naive prediction.

\section{Other decay modes}\label{U}

\begin{table*}[t!]
\caption{The number of independent $U$-spin amplitudes for different systems, in which $\textcircled{1}$, $\textcircled{2}$, $\textcircled{3}$, and $\textcircled{4}$ represent the $U$-spin singlet, doublet, triplet, and quartet, respectively.  }\label{y}
\begin{tabular}{|c|c|c|c|c|c|c|c|}\hline
 ~System~  &  ~$\textcircled{1}\to \textcircled{2}+\textcircled{2}$~ & ~$\textcircled{1}\to \textcircled{3}+\textcircled{3}$~ &  ~$\textcircled{1}\to \textcircled{1}+\textcircled{3}$~    & ~$\textcircled{1}\to \textcircled{2}+\textcircled{4}$~& ~$\textcircled{2}\to \textcircled{2}+\textcircled{3}$~ & ~$\textcircled{2}\to \textcircled{1}+\textcircled{2}$~& ~$\textcircled{2}\to \textcircled{1}+\textcircled{4}$~\\\hline
$N$ & $1$ & $1$ & $1$ & $1$ & $2$ & $1$ & $1$\\\hline
\end{tabular}
\end{table*}

The $U$-spin rate sum rules beyond the $U$-spin symmetry can be found systematically through a general procedure.
If a $U$-spin system receives contribution from only one $U$-spin amplitude in the $U$-spin limit, and contains two $U$-spin conjugate pairs, there must be a rate $U$-spin sum rule protected against first-order $U$-spin breaking in this system.
For example, the $U$-spin system \eqref{f4} is a $\textcircled{1}\to \textcircled{2}+\textcircled{2}$ system.
The number of independent $U$-spin amplitudes is one.
A rate $U$-spin sum rule \eqref{s7} is derived by summing the modulus squared of the amplitudes of the two $U$-spin conjugate pairs and equate the two sums to each other.
The number of independent $U$-spin amplitudes for various $U$-spin systems is summarized in Table~\ref{y}.
According to Table~\ref{y}, it is found the rate sum rule exists in most $U$-spin systems of charmed baryon decays, except for the $\textcircled{2}\to \textcircled{2}+\textcircled{3}$ system.

Following the above procedure, we find rate sum rules for singly and doubly charmed baryon decays that hold up to first-order $U$-spin breaking as follows:
\begin{align}\label{s8}
  |\mathcal{A}(\Xi^0_c\to \Sigma^+K^-)|^2+\frac{|\mathcal{A}(\Xi^0_c\to p\pi^-)|^2}{\tan^4\theta_C}=
  \frac{|\mathcal{A}(\Xi^0_c\to pK^-)|^2}{\tan^2\theta_C} +\frac{|\mathcal{A}(\Xi^0_c\to \Sigma^+\pi^-)|^2}{\tan^2\theta_C} +  \mathcal{O}(\epsilon^2),
\end{align}
\begin{align}
  3|\mathcal{A}(\Xi^0_c\to& \Xi^{*-}\pi^+)|^2+\frac{3|\mathcal{A}(\Xi^0_c\to \Sigma^{*-}K^+)|^2}{\tan^4\theta_C}=|\mathcal{A}(\Xi^0_c\to \Omega^{-}K^+)|^2+\frac{|\mathcal{A}(\Xi^0_c\to \Delta^{-}\pi^+)|^2}{\tan^4\theta_C}
  \nonumber\\ \qquad \qquad&=
  \frac{3}{4}\frac{|\mathcal{A}(\Xi^0_c\to\Sigma^{*-}\pi^+)|^2}{\tan^2\theta_C} +\frac{3}{4}\frac{|\mathcal{A}(\Xi^0_c\to \Xi^{*-}K^+)|^2}{\tan^2\theta_C} +  \mathcal{O}(\epsilon^2),
\end{align}
\begin{align}
  |\mathcal{A}(\Xi^0_c\to \Sigma^{*+}K^-)|^2+\frac{|\mathcal{A}(\Xi^0_c\to \Delta^+\pi^-)|^2}{\tan^4\theta_C}=
  \frac{|\mathcal{A}(\Xi^0_c\to \Delta^+K^-)|^2}{\tan^2\theta_C} +\frac{|\mathcal{A}(\Xi^0_c\to \Sigma^{*+}\pi^-)|^2}{\tan^2\theta_C} +  \mathcal{O}(\epsilon^2),
\end{align}
\begin{align}
  |\mathcal{A}(\Lambda^+_c\to \Delta^{++}K^-)|^2+\frac{|\mathcal{A}(\Xi^+_c\to \Delta^{++}\pi^-)|^2}{\tan^4\theta_C}=
  \frac{|\mathcal{A}(\Lambda^+_c\to \Delta^{++}\pi^-)|^2}{\tan^2\theta_C} +\frac{|\mathcal{A}(\Xi^+_c\to \Delta^{++}K^-)|^2}{\tan^2\theta_C} +  \mathcal{O}(\epsilon^2),
\end{align}
\begin{align}
  |\mathcal{A}(\Xi^{++}_{cc}\to \Xi^{+}_c\pi^+)|^2+\frac{|\mathcal{A}(\Xi^{++}_{cc}\to \Lambda^{+}_cK^+)|^2}{\tan^4\theta_C}=
  \frac{|\mathcal{A}(\Xi^{++}_{cc}\to \Xi^{+}_cK^+)|^2}{\tan^2\theta_C} +\frac{|\mathcal{A}(\Xi^{++}_{cc}\to \Lambda^{+}_c\pi^+)|^2}{\tan^2\theta_C} +  \mathcal{O}(\epsilon^2),
\end{align}
\begin{align}
  |\mathcal{A}(\Xi^{+}_{cc}\to \Xi^{0}_c\pi^+)|^2+\frac{|\mathcal{A}(\Omega^{+}_{cc}\to \Xi^{0}_cK^+)|^2}{\tan^4\theta_C}=
  \frac{|\mathcal{A}(\Xi^{+}_{cc}\to \Xi^{0}_cK^+)|^2}{\tan^2\theta_C} +\frac{|\mathcal{A}(\Omega^{+}_{cc}\to \Xi^{0}_c\pi^+)|^2}{\tan^2\theta_C} +  \mathcal{O}(\epsilon^2),
\end{align}
\begin{align}
  |\mathcal{A}(\Xi^{++}_{cc}\to \Xi^{*+}_c\pi^+)|^2+\frac{|\mathcal{A}(\Xi^{++}_{cc}\to \Sigma^{+}_cK^+)|^2}{\tan^4\theta_C}=
  \frac{|\mathcal{A}(\Xi^{++}_{cc}\to \Xi^{*+}_cK^+)|^2}{\tan^2\theta_C} +\frac{|\mathcal{A}(\Xi^{++}_{cc}\to \Sigma^{+}_c\pi^+)|^2}{\tan^2\theta_C} +  \mathcal{O}(\epsilon^2),
\end{align}
\begin{align}
  |\mathcal{A}(\Xi^{+}_{cc}\to \Sigma^{++}_cK^-)|^2+\frac{|\mathcal{A}(\Omega^{+}_{cc}\to \Sigma^{++}_c\pi^-)|^2}{\tan^4\theta_C}=
  \frac{|\mathcal{A}(\Xi^{+}_{cc}\to \Sigma^{++}_c\pi^-)|^2}{\tan^2\theta_C} +\frac{|\mathcal{A}(\Omega^{+}_{cc}\to \Sigma^{++}_cK^-)|^2}{\tan^2\theta_C} +  \mathcal{O}(\epsilon^2),
\end{align}
\begin{align}
  |\mathcal{A}(\Xi^{++}_{cc}\to \Sigma^{+}D^+)|^2+\frac{|\mathcal{A}(\Xi^{++}_{cc}\to pD^+_s)|^2}{\tan^4\theta_C}=
  \frac{|\mathcal{A}(\Xi^{++}_{cc}\to pD^+)|^2}{\tan^2\theta_C} +\frac{|\mathcal{A}(\Xi^{++}_{cc}\to \Sigma^{+}D^+_s)|^2}{\tan^2\theta_C} +  \mathcal{O}(\epsilon^2),
\end{align}
\begin{align}
  |\mathcal{A}(\Xi^{+}_{cc}\to \Sigma^{+}D^0)|^2+\frac{|\mathcal{A}(\Omega^{+}_{cc}\to pD^0)|^2}{\tan^4\theta_C}=
  \frac{|\mathcal{A}(\Xi^{+}_{cc}\to pD^0)|^2}{\tan^2\theta_C} +\frac{|\mathcal{A}(\Omega^{+}_{cc}\to \Sigma^{+}D^0)|^2}{\tan^2\theta_C} +  \mathcal{O}(\epsilon^2),
\end{align}
\begin{align}
  |\mathcal{A}(\Xi^{++}_{cc}\to \Sigma^{*+}D^+)|^2+\frac{|\mathcal{A}(\Xi^{++}_{cc}\to \Delta^+D^+_s)|^2}{\tan^4\theta_C}=
  \frac{|\mathcal{A}(\Xi^{++}_{cc}\to \Delta^+D^+)|^2}{\tan^2\theta_C} +\frac{|\mathcal{A}(\Xi^{++}_{cc}\to \Sigma^{*+}D^+_s)|^2}{\tan^2\theta_C} +  \mathcal{O}(\epsilon^2),
\end{align}
\begin{align}\label{s10}
  |\mathcal{A}(\Xi^{+}_{cc}\to \Sigma^{*+}D^0)|^2+\frac{|\mathcal{A}(\Omega^{+}_{cc}\to \Delta^+D^0)|^2}{\tan^4\theta_C}=
  \frac{|\mathcal{A}(\Xi^{+}_{cc}\to \Delta^+D^0)|^2}{\tan^2\theta_C} +\frac{|\mathcal{A}(\Omega^{+}_{cc}\to \Sigma^{*+}D^0)|^2}{\tan^2\theta_C} +  \mathcal{O}(\epsilon^2).
\end{align}
The $U$-spin amplitude decompositions of the above decay channels are presented in Appendix~\ref{amp}.
Note that the nonperturbative parameters $\epsilon_1$, $\epsilon_2$ ... and $a_{1,2}$, $a_{1,2}^\prime$ ... are distinct in different $U$-spin systems.
The LHS and RHS of Eqs.~\eqref{s8} $\sim$ \eqref{s10} can be replaced by the product of the modulus of the corresponding decay amplitudes like Eqs.~\eqref{s12} and \eqref{s13}.
And also, the decay parameters $\alpha$, $\beta$, $\gamma$ of these modes form the decay parameter sum rules like Eqs.~\eqref{s17}$ \sim $\eqref{s18}.
The derivation is similar to that presented in the last section.
Note that the Lee-Yang parameters $\alpha$, $\beta$, and $\gamma$ for the decuplet-baryon final states are defined as
\begin{align}
\alpha=\frac{2\mathcal{R}e (PD^*)}{|P|^2+ |D|^2},\qquad
\beta=\frac{2 \mathcal{I}m(PD^*)}{|P|^2+|D|^2},\qquad
\gamma=\frac{-|P|^2+|D|^2}{|P|^2+ |D|^2}.
\end{align}
Due to limited experimental data, we cannot test the validity of these $U$-spin relations derived in this work nor use them to predict unmeasured decay channels.
These sum rules are protected against first-order $U$-spin-breaking corrections and therefore provide relatively precise tests of the $U$-spin symmetry in the charmed baryon sector.

Compared with bottom-baryon decays, more relations protected against first-order $U$-spin-breaking corrections can be found in charmed baryon decays.
In bottom baryon decays, $U$-spin relations protected against first-order $U$-spin-breaking exist only in the $b\to c\overline ud/s$ and $b\to u\overline cd/s$ decays.
The decay modes dominated by the $b\to u\overline ud/s$ ($b\to c\overline cd/s$) transitions also receive contributions from the $b\to u\overline ud/s$ ($b\to c\overline cd/s$) transitions through $c$- ($u$-), and $t$-quark loop diagrams.
The products of CKM matrix elements associated with $c$- ($u$-) and $t$-quark loop diagrams differ from those associated with the tree diagrams in $b\to u\overline ud/s$ ($b\to c\overline cd/s$) decays and cannot be eliminated through flavor-symmetry cancellations.
The interference among contributions with different CKM structures prevents one from factoring out a common product of the CKM matrix elements.
In singly Cabibbo-suppressed charmed baryon decays, the products of the CKM matrix elements for the $c\to u\overline dd$ and $c\to u\overline ss$ transitions are $V_{cd}^*V_{ud}$ and $V_{cs}^*V_{us}$, respectively.
$V_{cd}^*V_{ud}$ and $V_{cs}^*V_{us}$ can be approximated as $V_{cd}^*V_{ud}\simeq -\cos\theta_C\sin\theta_C$ and $V_{cs}^*V_{us}\simeq \cos\theta_C\sin\theta_C$.
The contributions from b-quark loop diagrams are negligible because the CKM factor $V_{cb}^*V_{ub}$ is much smaller than $\cos\theta_C\sin\theta_C$.
Consequently, common CKM factors can be factored out in the Cabibbo-favored, singly Cabibbo-suppressed, and doubly Cabibbo-suppressed decays allowing all of these weak transitions to be used in constructing $U$-spin relations beyond the $U$-spin limit.
High order $U$-spin relations in bottom baryon decays only connect two distinct weak-transitions, $b\to c\overline ud$ and $b\to c\overline us$, or $b\to u\overline cd$ and $b\to u\overline cs$.
Only a limited number of $U$-spin relations can be found \cite{Wen:2025ibn}.
In contrast, the high order $U$-spin relations in charmed baryon decays connect four distinct weak-transitions, $c\to u\overline ds$, $c\to u\overline dd$, $c\to u\overline ss$, and $c\to u\overline sd$.
More higher-order $U$-spin relations can be found in charmed baryon decays.

At present, experimental measurements of the branching fractions of doubly charmed baryon decays remain highly challenging.
The $U$-spin relations presented in this work are derived within the narrow-width approximation.
For decay modes involving decuplet baryons in the final state, the final-state baryons generally have relatively large decay widths, especially the $\Delta$ baryons.
Finite-width effects, resonance line shapes, and interference effects in the corresponding multi-body final states may introduce additional uncertainties into these relations.
Compared to decays into decuplet baryons, charmed baryon decays into octet baryons provide cleaner tests of $U$-spin symmetry.
The main difficulty in testing the $U$-spin sum rules proposed in this work lines in measurements of the branching fractions of the doubly Cabibbo-suppressed modes.
The doubly Cabibbo-suppressed mode in Eq.~\eqref{s7}, $\Xi^0_c\to \Sigma^-K^+$, is
difficult to observe experimentally because the $\Sigma^-$ reconstructed through $\Sigma^-\to n\pi^-$.
As a result, Eq.~\eqref{s8} provides one of the most promising relations for testing $U$-spin symmetry since both final-state particles of the doubly Cabibbo-suppressed decay mode $\Xi^0_c\to p\pi^-$ are experimentally accessible.
The branching fraction of $\Xi^0_c\to p\pi^-$ is roughly estimated as $\mathcal{B}r(\Xi^0_c\to p\pi^-)\simeq \tan^4\theta_C\mathcal{B}r(\Xi^0_c\to \Sigma^+K^-)\approx4.5\times 10^{-6}$ \cite{PDG}.
The prospect of measuring $\mathcal{B}r(\Xi^0_c\to p\pi^-)$ at LHCb can be assessed using the reference channel $\Xi_c^0\to pK^-K^-\pi^+$.
Using the Run 1 data set corresponding to an integrated luminosity of $3~\mathrm{fb}^{-1}$, LHCb reconstructed $N(\Xi_c^0\to pK^-K^-\pi^+)=(21.6\pm0.2)\times10^{3}$ signal candidates \cite{LHCb:2019ldj}.
The absolute branching fraction of $\Xi_c^0\to pK^-K^-\pi^+$ given by Belle is
$\mathcal{B}(\Xi_c^0\to pK^-K^-\pi^+)=(5.8\pm2.3\pm0.5)\times10^{-3}$ \cite{Belle:2018kzz}.
Then we have $\mathcal{B}(\Xi_c^0\to p\pi^-)/\mathcal{B}(\Xi_c^0\to pK^-K^-\pi^+)\simeq7.8\times10^{-4}$.
If one naively assumes identical reconstruction efficiencies for the two decay modes, the expected signal yield in the Run 1 sample would be $N(p\pi^-)
\simeq 2.16\times10^{4} \times7.8\times10^{-4}\approx17$.
Assuming that the signal yield scales linearly with integrated luminosity and neglecting changes in detector performance, trigger efficiencies, and running conditions, Upgrade II of LHCb will collect a data set of $300~\mathrm{fb}^{-1}$, which could yield around $1700$ signal candidates of the $\Xi^0_c\to p\pi^-$ decay.

\section{Summary}\label{summary}

In this work, we studied $U$-spin sum rules for the two-body nonleptonic decays of charmed baryons including the first order $U$-spin breaking.
Several sum rules for decay rates and decay parameters that remain valid beyond the exact $U$-spin symmetry have been proposed.
These sum rules are protected against first-order $U$-spin breaking.
Their uncertainties are therefore expected to be smaller than the typical size of $U$-spin breaking.
Thus, these $U$-spin sum rules could provide relatively precise tests of $U$-spin symmetry in the charmed baryon decays.

\begin{acknowledgements}

This work was supported in part by Scientific Research Fund of Hunan Provincial Department under No. 25B0090.

\end{acknowledgements}

\begin{appendix}

\section{Angular momentum coupling rule}\label{amcr}

In this appendix, we derive the angular-momentum coupling relation in Eq.~\eqref{s3}.
The matrix element in Eq.~\eqref{s3} is assumed to represent a rotationally invariant coupling between two two-particle angular-momentum states.
The initial- and final-state angular momenta can be coupled separately to the same intermediate total angular momentum $J$ and its third component $M$.
The coupling coefficient can therefore be expanded as
\begin{align}
\mathcal{C}_{m_1m_2;m_3m_4} &\equiv \left\langle j_1,m_1;j_2,m_2 \middle| j_3,m_3;j_4,m_4 \right\rangle = \sum_{J,M} \mathcal{A}_{J}\, C^{JM}_{j_1m_1,j_2m_2} \left( C^{JM}_{j_3m_3,j_4m_4} \right)^{*}, \label{eq:coupling-expansion}
\end{align}
where
\begin{align}
C^{JM}_{j_1m_1,j_2m_2} \equiv \left\langle j_1,m_1;j_2,m_2 \middle| J,M \right\rangle \label{eq:cg-definition}
\end{align}
is a Clebsch-Gordan coefficient.
$\mathcal{A}_{J}$ is a reduced coupling coefficient independent of the magnetic quantum numbers. In the Condon--Shortley phase convention, the Clebsch--Gordan coefficients satisfy
\begin{align}
C^{J,-M}_{j_1,-m_1;j_2,-m_2} = (-1)^{j_1+j_2-J} C^{JM}_{j_1m_1,j_2m_2}. \label{eq:cg-reversal-initial}
\end{align}
Similarly, for the final-state angular momenta,
\begin{align}
C^{J,-M}_{j_3,-m_3;j_4,-m_4} = (-1)^{j_3+j_4-J} C^{JM}_{j_3m_3,j_4m_4}. \label{eq:cg-reversal-final}
\end{align}
After reversing all magnetic quantum numbers, the coupling coefficient becomes \begin{align}
\mathcal{C}_{-m_1,-m_2;-m_3,-m_4} &= \sum_{J,M} \mathcal{A}_{J}\, C^{J,-M}_{j_1,-m_1;j_2,-m_2} \left( C^{J,-M}_{j_3,-m_3;j_4,-m_4} \right)^{*}. \label{eq:reversed-expansion}
\end{align}
Using Eqs.~\eqref{eq:cg-reversal-initial} and \eqref{eq:cg-reversal-final}, one obtains
\begin{align}
\mathcal{C}_{-m_1,-m_2;-m_3,-m_4} &= \sum_{J,M} \mathcal{A}_{J}\, (-1)^{j_1+j_2-J} (-1)^{j_3+j_4-J} C^{JM}_{j_1m_1,j_2m_2} \left( C^{JM}_{j_3m_3,j_4m_4} \right)^{*} \nonumber\\ &\quad= \sum_{J,M} \mathcal{A}_{J}\, (-1)^{j_1+j_2+j_3+j_4-2J} C^{JM}_{j_1m_1,j_2m_2} \left( C^{JM}_{j_3m_3,j_4m_4} \right)^{*}. \label{eq:combined-phase}
\end{align}
For an allowed coupling of the final-state angular momenta, we have $j_3+j_4-J\in\mathbb{Z}$.
Then $2\left(j_3+j_4-J\right)$ is an even integer.
The exponent in Eq.~\eqref{eq:combined-phase} can be rearranged as
\begin{align} j_1+j_2+j_3+j_4-2J &= j_1+j_2-j_3-j_4 + 2\left(j_3+j_4-J\right). \label{eq:phase-rearrangement}
\end{align}
Since $2\left(j_3+j_4-J\right)$ is an even integer, it follows that
\begin{align}
(-1)^{j_1+j_2+j_3+j_4-2J} = (-1)^{j_1+j_2-j_3-j_4}. \label{eq:phase-equivalence}
\end{align}
Substituting Eq.~\eqref{eq:phase-equivalence} into Eq.~\eqref{eq:combined-phase}, we obtain
\begin{align}
\left\langle j_1,-m_1;j_2,-m_2 \middle| j_3,-m_3;j_4,-m_4 \right\rangle = (-1)^{j_1+j_2-j_3-j_4} \left\langle j_1,m_1;j_2,m_2 \middle| j_3,m_3;j_4,m_4 \right\rangle. \label{eq:final-coupling-reversal}
\end{align}

\section{$U$-spin amplitude decompositions}\label{amp}

In this appendix, we list the $U$-spin amplitudes for the $U$-spin systems that are analyzed in Eqs.~\eqref{s8} $\sim$ \eqref{s10}.

\begin{enumerate}
  \item The $U$-spin amplitudes for the $U$-spin system
\begin{align}
    \begin{array}{c}
      \Xi^0_c
    \end{array}
  \to
    \left(
    \begin{array}{c}
      \pi^- \\
      K^- \\
    \end{array}
  \right)
    \left(
    \begin{array}{c}
      p \\
     \Sigma^+ \\
    \end{array}
  \right)
\end{align}
are expressed as
\begin{align}
 \mathcal{A}(\Xi^0_c\to \Sigma^+K^-) &= -\cos^2\theta_C\mathcal{A}_{1}(1- \epsilon_1+a_1\epsilon^2_1-a^\prime_1\epsilon^3_1+...),\nonumber\\
\mathcal{A}(\Xi^0_c\to p\pi^-) &= -\sin^2\theta_C\mathcal{A}_{1}(1+ \epsilon_1+a_1\epsilon^2_1+a^\prime_1\epsilon^3_1+...),\nonumber\\
 \mathcal{A}(\Xi^0_c\to pK^-)&= \cos\theta_C\sin\theta_C\,\mathcal{A}_{1}(1- \epsilon_2+a_2\epsilon^2_2-a^\prime_2\epsilon^3_2+...),\nonumber\\ \mathcal{A}(\Xi^0_c\to \Sigma^+\pi^-)&= \cos\theta_C\sin\theta_C\,\mathcal{A}_{1}(1+ \epsilon_2+a_2\epsilon^2_2+a^\prime_2\epsilon^3_2+...).
\end{align}
  \item The $U$-spin amplitudes for the $U$-spin system
\begin{align}
    \begin{array}{c}
      \Xi^0_c
    \end{array}
  \to
    \left(
    \begin{array}{c}
      K^+ \\
      \pi^+ \\
    \end{array}
  \right)
    \left(
    \begin{array}{c}
    \Delta^{-} \\
      \Sigma^{*-} \\
     \Xi^{*-} \\
     \Omega^{-}\\
    \end{array}
  \right)
\end{align}
are expressed as
\begin{align}
 \mathcal{A}(\Xi^0_c\to \Xi^{*-}\pi^+) &= -\frac{1}{2}\cos^2\theta_C\mathcal{A}_{1}(1- \epsilon_1+a_1\epsilon^2_1-a^\prime_1\epsilon^3_1+...),\nonumber\\
\mathcal{A}(\Xi^0_c\to \Sigma^{*-}K^+) &= \frac{1}{2}\sin^2\theta_C\mathcal{A}_{1}(1+ \epsilon_1+a_1\epsilon^2_1+a^\prime_1\epsilon^3_1+...),\nonumber\\
 \mathcal{A}(\Xi^0_c\to \Omega^{-}K^+)&= \frac{\sqrt{3}}{2}\cos^2\theta_C\,\mathcal{A}_{1}(1- \epsilon_2+a_2\epsilon^2_2-a^\prime_2\epsilon^3_2+...),\nonumber\\ \mathcal{A}(\Xi^0_c\to \Delta^{-}\pi^+)&= -\frac{\sqrt{3}}{2}\sin^2\theta_C\,\mathcal{A}_{1}(1+ \epsilon_2+a_2\epsilon^2_2+a^\prime_2\epsilon^3_2+...),\nonumber\\
 \mathcal{A}(\Xi^0_c\to \Sigma^{*-}\pi^+)&= \cos\theta_C\sin\theta_C\,\mathcal{A}_{1}(1- \epsilon_3+a_3\epsilon^2_3-a^\prime_3\epsilon^3_3+...),\nonumber\\
  \mathcal{A}(\Xi^0_c\to \Xi^{*-}K^+)&= -\cos\theta_C\sin\theta_C\,\mathcal{A}_{1}(1+ \epsilon_3+a_3\epsilon^2_3+a^\prime_3\epsilon^3_3+...).
\end{align}
  \item The $U$-spin amplitudes for the $U$-spin system
\begin{align}
    \begin{array}{c}
      \Xi^0_c
    \end{array}
  \to
    \left(
    \begin{array}{c}
    \pi^- \\
      K^- \\
    \end{array}
  \right)
    \left(
    \begin{array}{c}
    \Delta^{+} \\
      \Sigma^{*+} \\
    \end{array}
  \right)
\end{align}
are expressed as
\begin{align}
 \mathcal{A}(\Xi^0_c\to \Sigma^{*+}K^-) &= -\cos^2\theta_C\mathcal{A}_{1}(1- \epsilon_1+a_1\epsilon^2_1-a^\prime_1\epsilon^3_1+...),\nonumber\\
\mathcal{A}(\Xi^0_c\to \Delta^{+}\pi^-) &= -\sin^2\theta_C\mathcal{A}_{1}(1+ \epsilon_1+a_1\epsilon^2_1+a^\prime_1\epsilon^3_1+...),\nonumber\\
 \mathcal{A}(\Xi^0_c\to \Delta^{+}K^-)&= \cos\theta_C\sin\theta_C\,\mathcal{A}_{1}(1- \epsilon_2+a_2\epsilon^2_2-a^\prime_2\epsilon^3_2+...),\nonumber\\
  \mathcal{A}(\Xi^0_c\to \Sigma^{*+}\pi^-)&= \cos\theta_C\sin\theta_C\,\mathcal{A}_{1}(1+ \epsilon_2+a_2\epsilon^2_2+a^\prime_2\epsilon^3_2+...).
\end{align}
  \item The $U$-spin amplitudes for the $U$-spin system
\begin{align}
   \left(  \begin{array}{c}
     \Lambda^+_c\\
      \Xi^+_c \\
    \end{array} \right)
  \to
    \left(
    \begin{array}{c}
    \pi^- \\
      K^- \\
    \end{array}
  \right)\,
   \Delta^{++}
\end{align}
are expressed as
\begin{align}
 \mathcal{A}(\Lambda^+_c\to \Delta^{++}K^-) &= -\sqrt{\frac{2}{3}}\cos^2\theta_C\mathcal{A}_{1/2}(1- \epsilon_1+a_1\epsilon^2_1-a^\prime_1\epsilon^3_1+...),\nonumber\\
\mathcal{A}(\Xi^+_c\to \Delta^{++}\pi^-) &= \sqrt{\frac{2}{3}}\sin^2\theta_C\mathcal{A}_{1/2}(1+ \epsilon_1+a_1\epsilon^2_1+a^\prime_1\epsilon^3_1+...),\nonumber\\
 \mathcal{A}(\Lambda^+_c\to \Delta^{++}\pi^-)&= \sqrt{\frac{2}{3}}\cos\theta_C\sin\theta_C\,\mathcal{A}_{1/2}(1- \epsilon_2+a_2\epsilon^2_2-a^\prime_2\epsilon^3_2+...),\nonumber\\
  \mathcal{A}(\Xi^+_c\to \Delta^{++}K^-)&= -\sqrt{\frac{2}{3}}\cos\theta_C\sin\theta_C\,\mathcal{A}_{1/2}(1+ \epsilon_2+a_2\epsilon^2_2+a^\prime_2\epsilon^3_2+...).
\end{align}
  \item The $U$-spin amplitudes for the $U$-spin system
\begin{align}
    \begin{array}{c}
      \Xi^{++}_{cc}
    \end{array}
  \to
    \left(
    \begin{array}{c}
      K^+ \\
      \pi^+ \\
    \end{array}
  \right)
    \left(
    \begin{array}{c}
      \Lambda^+_c \\
    \Xi^+_c \\
    \end{array}
  \right)
\end{align}
are expressed as
\begin{align}
 \mathcal{A}(\Xi^{++}_{cc}\to \Xi^+_c\pi^+) &= -\cos^2\theta_C\mathcal{A}_{1}(1- \epsilon_1+a_1\epsilon^2_1-a^\prime_1\epsilon^3_1+...),\nonumber\\
\mathcal{A}(\Xi^{++}_{cc}\to \Lambda^+_cK^+) &= -\sin^2\theta_C\mathcal{A}_{1}(1+ \epsilon_1+a_1\epsilon^2_1+a^\prime_1\epsilon^3_1+...),\nonumber\\
 \mathcal{A}(\Xi^{++}_{cc}\to \Xi^+_cK^+)&= \cos\theta_C\sin\theta_C\,\mathcal{A}_{1}(1- \epsilon_2+a_2\epsilon^2_2-a^\prime_2\epsilon^3_2+...),\nonumber\\ \mathcal{A}(\Xi^{++}_{cc}\to \Lambda^+_c\pi^+)&= \cos\theta_C\sin\theta_C\,\mathcal{A}_{1}(1+ \epsilon_2+a_2\epsilon^2_2+a^\prime_2\epsilon^3_2+...).
\end{align}
  \item The $U$-spin amplitudes for the $U$-spin system
\begin{align}
   \left(  \begin{array}{c}
     \Xi^+_{cc}\\
      \Omega^+_{cc} \\
    \end{array} \right)
  \to
    \left(
    \begin{array}{c}
    K^+ \\
      \pi^+ \\
    \end{array}
  \right)\,
   \Xi^{0}_c
\end{align}
are expressed as
\begin{align}
 \mathcal{A}(\Xi^{+}_{cc}\to \Xi^{0}_c\pi^+) &= -\sqrt{\frac{2}{3}}\cos^2\theta_C\mathcal{A}_{1/2}(1- \epsilon_1+a_1\epsilon^2_1-a^\prime_1\epsilon^3_1+...),\nonumber\\
\mathcal{A}(\Omega^{+}_{cc}\to \Xi^{0}_cK^+) &= \sqrt{\frac{2}{3}}\sin^2\theta_C\mathcal{A}_{1/2}(1+ \epsilon_1+a_1\epsilon^2_1+a^\prime_1\epsilon^3_1+...),\nonumber\\
 \mathcal{A}(\Xi^{+}_{cc}\to \Xi^{0}_cK^+)&= \sqrt{\frac{2}{3}}\cos\theta_C\sin\theta_C\,\mathcal{A}_{1/2}(1- \epsilon_2+a_2\epsilon^2_2-a^\prime_2\epsilon^3_2+...),\nonumber\\
  \mathcal{A}(\Omega^{+}_{cc}\to \Xi^{0}_c\pi^+)&= -\sqrt{\frac{2}{3}}\cos\theta_C\sin\theta_C\,\mathcal{A}_{1/2}(1+ \epsilon_2+a_2\epsilon^2_2+a^\prime_2\epsilon^3_2+...).
\end{align}
  \item The $U$-spin amplitudes for the $U$-spin system
\begin{align}
    \begin{array}{c}
      \Xi^{++}_{cc}
    \end{array}
  \to
    \left(
    \begin{array}{c}
      K^+ \\
      \pi^+ \\
    \end{array}
  \right)
    \left(
    \begin{array}{c}
      \Sigma^+_c \\
    \Xi^{*+}_c \\
    \end{array}
  \right)
\end{align}
are expressed as
\begin{align}
 \mathcal{A}(\Xi^{++}_{cc}\to \Xi^{*+}_c\pi^+) &= -\cos^2\theta_C\mathcal{A}_{1}(1- \epsilon_1+a_1\epsilon^2_1-a^\prime_1\epsilon^3_1+...),\nonumber\\
\mathcal{A}(\Xi^{++}_{cc}\to \Sigma^+_cK^+) &= -\sin^2\theta_C\mathcal{A}_{1}(1+ \epsilon_1+a_1\epsilon^2_1+a^\prime_1\epsilon^3_1+...),\nonumber\\
 \mathcal{A}(\Xi^{++}_{cc}\to \Xi^{*+}_cK^+)&= \cos\theta_C\sin\theta_C\,\mathcal{A}_{1}(1- \epsilon_2+a_2\epsilon^2_2-a^\prime_2\epsilon^3_2+...),\nonumber\\ \mathcal{A}(\Xi^{++}_{cc}\to \Sigma^+_c\pi^+)&= \cos\theta_C\sin\theta_C\,\mathcal{A}_{1}(1+ \epsilon_2+a_2\epsilon^2_2+a^\prime_2\epsilon^3_2+...).
\end{align}
  \item The $U$-spin amplitudes for the $U$-spin system
\begin{align}
   \left(  \begin{array}{c}
     \Xi^+_{cc}\\
      \Omega^+_{cc} \\
    \end{array} \right)
  \to
    \left(
    \begin{array}{c}
    \pi^- \\
      K^- \\
    \end{array}
  \right)\,
   \Sigma^{++}_c
\end{align}
are expressed as
\begin{align}
 \mathcal{A}(\Xi^{+}_{cc}\to \Sigma^{++}_cK^-) &= -\sqrt{\frac{2}{3}}\cos^2\theta_C\mathcal{A}_{1/2}(1- \epsilon_1+a_1\epsilon^2_1-a^\prime_1\epsilon^3_1+...),\nonumber\\
\mathcal{A}(\Omega^{+}_{cc}\to \Sigma^{++}_c\pi^-) &= \sqrt{\frac{2}{3}}\sin^2\theta_C\mathcal{A}_{1/2}(1+ \epsilon_1+a_1\epsilon^2_1+a^\prime_1\epsilon^3_1+...),\nonumber\\
 \mathcal{A}(\Xi^{+}_{cc}\to \Sigma^{++}_c\pi^-)&= \sqrt{\frac{2}{3}}\cos\theta_C\sin\theta_C\,\mathcal{A}_{1/2}(1- \epsilon_2+a_2\epsilon^2_2-a^\prime_2\epsilon^3_2+...),\nonumber\\
  \mathcal{A}(\Omega^{+}_{cc}\to \Sigma^{++}_cK^-)&= -\sqrt{\frac{2}{3}}\cos\theta_C\sin\theta_C\,\mathcal{A}_{1/2}(1+ \epsilon_2+a_2\epsilon^2_2+a^\prime_2\epsilon^3_2+...).
\end{align}
  \item The $U$-spin amplitudes for the $U$-spin system
\begin{align}
    \begin{array}{c}
      \Xi^{++}_{cc}
    \end{array}
  \to
    \left(
    \begin{array}{c}
      D^+_s \\
      D^+ \\
    \end{array}
  \right)
    \left(
    \begin{array}{c}
      p \\
    \Sigma^+ \\
    \end{array}
  \right)
\end{align}
are expressed as
\begin{align}
 \mathcal{A}(\Xi^{++}_{cc}\to \Sigma^{+}D^+) &= -\cos^2\theta_C\mathcal{A}_{1}(1- \epsilon_1+a_1\epsilon^2_1-a^\prime_1\epsilon^3_1+...),\nonumber\\
\mathcal{A}(\Xi^{++}_{cc}\to pD^+_s) &= -\sin^2\theta_C\mathcal{A}_{1}(1+ \epsilon_1+a_1\epsilon^2_1+a^\prime_1\epsilon^3_1+...),\nonumber\\
 \mathcal{A}(\Xi^{++}_{cc}\to pD^+)&= \cos\theta_C\sin\theta_C\,\mathcal{A}_{1}(1- \epsilon_2+a_2\epsilon^2_2-a^\prime_2\epsilon^3_2+...),\nonumber\\ \mathcal{A}(\Xi^{++}_{cc}\to \Sigma^+D^+_s)&= \cos\theta_C\sin\theta_C\,\mathcal{A}_{1}(1+ \epsilon_2+a_2\epsilon^2_2+a^\prime_2\epsilon^3_2+...).
\end{align}
  \item The $U$-spin amplitudes for the $U$-spin system
\begin{align}
   \left(  \begin{array}{c}
     \Xi^+_{cc}\\
      \Omega^+_{cc} \\
    \end{array} \right)
  \to D^0
    \left(
    \begin{array}{c}
    p \\
     \Sigma^+ \\
    \end{array}
  \right)
\end{align}
are expressed as
\begin{align}
 \mathcal{A}(\Xi^{+}_{cc}\to \Sigma^{+}D^0) &= -\sqrt{\frac{2}{3}}\cos^2\theta_C\mathcal{A}_{1/2}(1- \epsilon_1+a_1\epsilon^2_1-a^\prime_1\epsilon^3_1+...),\nonumber\\
\mathcal{A}(\Omega^{+}_{cc}\to pD^0) &= \sqrt{\frac{2}{3}}\sin^2\theta_C\mathcal{A}_{1/2}(1+ \epsilon_1+a_1\epsilon^2_1+a^\prime_1\epsilon^3_1+...),\nonumber\\
 \mathcal{A}(\Xi^{+}_{cc}\to pD^0)&= \sqrt{\frac{2}{3}}\cos\theta_C\sin\theta_C\,\mathcal{A}_{1/2}(1- \epsilon_2+a_2\epsilon^2_2-a^\prime_2\epsilon^3_2+...),\nonumber\\
  \mathcal{A}(\Omega^{+}_{cc}\to \Sigma^{+}D^0)&= -\sqrt{\frac{2}{3}}\cos\theta_C\sin\theta_C\,\mathcal{A}_{1/2}(1+ \epsilon_2+a_2\epsilon^2_2+a^\prime_2\epsilon^3_2+...).
\end{align}
  \item The $U$-spin amplitudes for the $U$-spin system
\begin{align}
    \begin{array}{c}
      \Xi^{++}_{cc}
    \end{array}
  \to
    \left(
    \begin{array}{c}
      D^+_s \\
      D^+ \\
    \end{array}
  \right)
    \left(
    \begin{array}{c}
      \Delta^+ \\
    \Sigma^{*+} \\
    \end{array}
  \right)
\end{align}
are expressed as
\begin{align}
 \mathcal{A}(\Xi^{++}_{cc}\to \Sigma^{*+}D^+) &= -\cos^2\theta_C\mathcal{A}_{1}(1- \epsilon_1+a_1\epsilon^2_1-a^\prime_1\epsilon^3_1+...),\nonumber\\
\mathcal{A}(\Xi^{++}_{cc}\to \Delta^+D^+_s) &= -\sin^2\theta_C\mathcal{A}_{1}(1+ \epsilon_1+a_1\epsilon^2_1+a^\prime_1\epsilon^3_1+...),\nonumber\\
 \mathcal{A}(\Xi^{++}_{cc}\to \Delta^+D^+)&= \cos\theta_C\sin\theta_C\,\mathcal{A}_{1}(1- \epsilon_2+a_2\epsilon^2_2-a^\prime_2\epsilon^3_2+...),\nonumber\\ \mathcal{A}(\Xi^{++}_{cc}\to \Sigma^{*+}D^+_s)&= \cos\theta_C\sin\theta_C\,\mathcal{A}_{1}(1+ \epsilon_2+a_2\epsilon^2_2+a^\prime_2\epsilon^3_2+...).
\end{align}
  \item The $U$-spin amplitudes for the $U$-spin system
\begin{align}
   \left(  \begin{array}{c}
     \Xi^+_{cc}\\
      \Omega^+_{cc} \\
    \end{array} \right)
  \to D^0
    \left(
    \begin{array}{c}
    \Delta^+ \\
     \Sigma^{*+} \\
    \end{array}
  \right)
\end{align}
are expressed as
\begin{align}
 \mathcal{A}(\Xi^{+}_{cc}\to \Sigma^{*+}D^0) &= -\sqrt{\frac{2}{3}}\cos^2\theta_C\mathcal{A}_{1/2}(1- \epsilon_1+a_1\epsilon^2_1-a^\prime_1\epsilon^3_1+...),\nonumber\\
\mathcal{A}(\Omega^{+}_{cc}\to \Delta^+D^0) &= \sqrt{\frac{2}{3}}\sin^2\theta_C\mathcal{A}_{1/2}(1+ \epsilon_1+a_1\epsilon^2_1+a^\prime_1\epsilon^3_1+...),\nonumber\\
 \mathcal{A}(\Xi^{+}_{cc}\to \Delta^+D^0)&= \sqrt{\frac{2}{3}}\cos\theta_C\sin\theta_C\,\mathcal{A}_{1/2}(1- \epsilon_2+a_2\epsilon^2_2-a^\prime_2\epsilon^3_2+...),\nonumber\\
  \mathcal{A}(\Omega^{+}_{cc}\to \Sigma^{*+}D^0)&= -\sqrt{\frac{2}{3}}\cos\theta_C\sin\theta_C\,\mathcal{A}_{1/2}(1+ \epsilon_2+a_2\epsilon^2_2+a^\prime_2\epsilon^3_2+...).
\end{align}
\end{enumerate}

\end{appendix}

\end{document}